\documentstyle[preprint,aps]{revtex}
\newtheorem{emaxiom}{Electromagnetic Axiom}
\newtheorem{gaxiom}{Gravitational Axiom}
\begin{document}

\title{An axiomatic approach to electromagnetic and gravitational
radiation reaction of particles in curved spacetime}

\author{Theodore C. Quinn and Robert M. Wald}

\address{Enrico Fermi Institute and Department of Physics\\
         University of Chicago\\
         5640 S. Ellis Avenue\\
         Chicago, Illinois 60637-1433}

\date{}

\maketitle

\begin{abstract}
The problem of determining the electromagnetic and gravitational
``self-force'' on a particle in a curved spacetime is investigated
using an axiomatic approach. In the electromagnetic case, our key
postulate is a ``comparison axiom'', which states that whenever two
particles of the same charge $e$ have the same magnitude of
acceleration, the difference in their self-force is given by the
ordinary Lorentz force of the difference in their (suitably compared)
electromagnetic fields. We thereby derive an expression for the
electromagnetic self-force which agrees with that of DeWitt and Brehme
as corrected by Hobbs. Despite several important differences, our
analysis of the gravitational self-force proceeds in close parallel
with the electromagnetic case. In the gravitational case, our final
expression for the (reduced order) equations of motion shows that the
deviation from geodesic motion arises entirely from a ``tail term'',
in agreement with recent results of Mino et al. Throughout the paper,
we take the view that ``point particles'' do not make sense as
fundamental objects, but that ``point particle equations of motion''
do make sense as means of encoding information about the motion of an
extended body in the limit where not only the size but also the charge
and mass of the body go to zero at a suitable rate. Plausibility
arguments for the validity of our comparison axiom are given by
considering the limiting behavior of the self-force on extended
bodies.
\end{abstract}


\section{Introduction}

In this paper, we shall investigate the motion of an isolated body
coupled to classical fields in the limit where the spatial extent of
the body is small enough that the detailed structure of the body is
unimportant, but where the lowest order effects of the ``self-field''
of the body (which are responsible for ``radiation reaction'') are
taken into account. Specifically, we shall consider (i) the motion of
a charged body coupled to a Maxwell field on an arbitrary, fixed
curved background and (ii) the motion of a massive body in an
otherwise vacuum spacetime in general relativity.

No difficulty of principle is encountered in the calculation of the
motion of any extended body once one has specified what matter fields
compose the body and the equations of motion of these matter
fields. If one then gives the initial data for these matter fields (as
well as for the classical fields to which they are coupled), the
complete motion of the body is determined unambiguously by the full
set of continuum field equations. However, in practice, the details of
the motion of an extended body will be very complicated and will
depend on the detailed ``internal structure'' of the body. Thus, if
one deals with general, extended bodies, it is highly unlikely that
any simple results can be obtained which apply to large classes of
systems.

One would expect the details of the internal structure of the body to
become less and less important in the limit as one makes the body
smaller. Thus, one obvious way to seek a class of simple, general
results is to take the limit of the continuum equations as the spatial
extent of the body goes to zero, thereby obtaining equations of motion
for a ``point particle'' idealization of an extended body. However, as
is well known, serious difficulties arise in attempting to take the
point particle limit in a straightforward manner, keeping the total
charge, $e$, and mass, $m$, of the body fixed. In the electromagnetic
case, the linearity of Maxwell's equations allows one to make sense of
the electromagnetic field of a body of finite charge in the point
particle limit. However, the stress-energy of the electromagnetic
field becomes singular in this limit, and the the stress-energy of the
matter fields comprising the body must become correspondingly singular
in order for it to ``hold itself together''.  Hence, there is no
reason to expect that a well defined point particle limit will
exist. In general relativity, the situation is even worse, since the
nonlinearity of Einstein's equation does not allow one even to make
sense of the ``gravitational field'' of a point mass \cite{gt};
physically, an extended body would presumably collapse to a black hole
before a point particle limit could be achieved. Thus, it does not
appear to be mathematically or physically sensible to attempt to take
a point particle limit---holding the charge or mass fixed---in either
the electromagnetic or gravitational cases.

Nevertheless, we believe that there should exist some simple and
general results regarding the motion of bodies in the limit where the
size of the body is sufficiently small (as compared, in particular,
with the scale of variation of the background electromagnetic field
and/or the radius of curvature of the background spacetime) {\em and}
the charge and mass of the body also are sufficiently small that
``self-field'' effects do not become dominant---but are not
negligible.  Mathematically rigorous results of this sort presumably
would take the form of statements about limits of smooth,
one-parameter families of solutions to the full continuum equations in
which both the spatial extent of the body and the total charge, $e$,
and/or mass, $m$, of the body are simultaneously taken to zero in a
suitable manner.  Geodesic motion should result in such a
limit.\footnote{This should follow directly from the theorem of Geroch
and Jang \cite{gj} that any world line in a fixed background spacetime
having the property that every neighborhood of it admits a conserved
stress energy tensor satisfying the dominant energy condition must be
a timelike geodesic.} For a charged body, the Lorentz force law in the
background electromagnetic field should then arise as a correction to
geodesic motion, with the acceleration of the body being of order
$e/m$. For such a charged body, we seek in this paper to find the
further corrections to this result to order $e^2/m$. Similarly, for an
uncharged body, we seek to obtain the corrections to geodesic motion
to order $m$ in the acceleration of the body.  These corrections arise
from the ``self-field'' of the body and usually are referred to as
``radiation reaction'' or ``self-force'' effects. In both cases, we
shall not seek to obtain any additional corrections to the motion due
to the finite size and asphericity of the bodies. There is an
extensive literature on the equations of motion of extended bodies
(see, e.g., Dixon \cite{dixon74} and references cited therein) which
takes into account the lowest order deviations from Lorentz force or
geodesic motion due to such finite size effects, but neglects the
$O(e^2/m)$ and $O(m)$ corrections which concern us here. Presumably,
to lowest order, the combined effects of both types of corrections
would be obtained by simply adding together the radiation reaction
force due to self-field effects given here and ``multipole'' forces
due to the background field which can be found elsewhere in the
literature.

Thus, the goal of this paper is to obtain effective equations of
motion for a point particle which are accurate to the orders specified
in the previous paragraph.  It should be emphasized that (in contrast
to some other analyses of radiation reaction phenomena) our philosophy
is {\em not} to view a point particle of finite charge or mass as a
fundamental object, but rather to view the point particle equations of
motion we obtain as a formal device to express approximate results for
the continuum theory, in the limit where not only the size but also
the charge and/or mass of the body are sufficiently small.

As already indicated above, it should be possible to derive the
results we seek rigorously, without any additional hypotheses, by
considering suitable one-parameter families of solutions in which the
size, charge, and mass of the extended body go to zero in an
appropriate manner, and suitable assumptions are made about the
composition and initial state of the body so that its deviations from
sphericity are kept under good control in this limit. However,
although we shall take a few first steps in this direction in
section~\ref{motivation}, we shall not proceed in this manner
here. Rather, we shall instead attempt to ``guess'' the correct
equations by defining a set of axioms which we believe the total force
on the particle should obey. These axioms uniquely determine the
force, and we then shall give an explicit prescription which satisfies
our axioms.

Our approach bears some similarity to an approach of Penrose
\cite{pen} for obtaining the electromagnetic self-force in the sense
that, in both approaches, the self-force is given by the ordinary
Lorentz force associated with a regularized electromagnetic
field. However, the regularization is accomplished in very different
ways in the two approaches. In the approach of \cite{pen}, the
electromagnetic field of a point particle is regularized by means of
the Kirchhoff-d'Adhemar formula, whereas in our approach, the
electromagnetic field is effectively regularized by considering
differences in the fields associated with different particle
trajectories in (possibly different) spacetimes. We have not
investigated whether these regularization schemes are equivalent.

In Section~\ref{sec:em}, we apply our approach to the case of charged
point particles coupled to Maxwell fields on an arbitrary curved
background. Our results agree with those of DeWitt and Brehme
\cite{db} (as corrected by Hobbs \cite{hobbs}), but the calculations
required to obtain the radiation reaction force are considerably
simpler and they generalize much more naturally to the gravitational
case. In Section~\ref{sec:gr}, we carry out this generalization to
obtain the $O(m)$ correction to geodesic motion for a point particle
propagating on an otherwise vacuum background. Our final results agree
with those recently obtained by Mino et al. \cite{mino}, but, again,
the calculations required in our approach are considerably simpler.

\section{Electromagnetic radiation reaction}
\label{sec:em}

\subsection{Introduction and motivation for the ``comparison axiom''}
\label{motivation} 

In this section, we shall consider an arbitrary spacetime~$(M,g_{ab})$
containing an arbitrary timelike world line~$z(\tau)$ with tangent
$u^b$ representing a point charge of charge~$e$ and mass~$m$. We
consider an arbitrary solution, $F^{ab}$, of Maxwell's equations with
the point particle as its source
\begin{equation}
\nabla_a F^{ab} = - 4 \pi e \int \delta(x,z(\tau))\,u^b(\tau)\,d\tau
\label{Ftotal}
\end{equation}
such that $F^{ab}$ is smooth except on the world line of the particle.
Our aim is to give a prescription for assigning to each point of the
world line a vector~$f^a$---which we shall refer to as the {\em total
electromagnetic force}---such that the point particle equation of
motion~$a^b = f^b/m$, will be valid to order $e^2/m$ as an equation of
motion for a sufficiently small, nearly spherical charged body, as
discussed in the previous section.

In order to motivate the axiomatic approach that we shall ultimately
adopt, let us see what happens if we attempt to derive a formula for
$f^a$ by considering a small, nearly spherical extended body and
taking a ``point particle limit''. To avoid unnecessary complications,
we shall initially restrict attention in the discussion below to the
case of a body moving in Minkowski spacetime.

The first difficult issue we encounter in this program is that of
obtaining a ``representative world line'' in the extended body which
can be viewed as describing the motion of the body, so that we may
contemplate the limiting behavior of this world line as the body
shrinks toward zero spatial extent. We shall not attempt to analyze
this issue here, but will merely assume that by methods similar to
those of Beiglb\"{o}ck \cite{beig}, it is possible to define a
``center of mass'' or ``center of motion'' world line such that---to
an excellent approximation when the body is sufficiently small and
nearly spherical---we have, at each point of the world line,
\begin{equation}
p^a = mu^a
\label{pmv}
\end{equation}
where $u^a$ is the tangent to the representative world line and $p^a$
is given by
\begin{eqnarray}
p^a &=& \int_{\Sigma} (T_{\rm body}^{ab}) \epsilon_{bcde},
\label{defp}
\end{eqnarray}
where $T_{\rm body}^{ab}$ is the stress-energy of the body ({\em not}
including the electromagnetic field) and the integral is taken over
the hyperplane $\Sigma$ perpendicular to $u^a$.  Here
$\epsilon_{abcd}$ denotes the volume four-form determined by the
metric, and the integrand is to be viewed as a vector-valued
three-form; the global parallelism of flat spacetime is used here to
define the integrals of tensor fields.  It should be noted that the
difficulties in controlling errors in eq.~(\ref{pmv}) and the
corresponding equation in curved spacetime probably constitute the
most serious obstacle to converting the arguments given in this paper
into rigorous theorems about radiation reaction forces which do not
rely on any additional axioms.

We define the force on the extended body by
\begin{equation}
f^a = u^b \nabla_b (p^a).
\label{deff}
\end{equation}
Given eq.~(\ref{pmv}), we that $a^a \equiv u^b \nabla_b u^a$ will be
given by the projection of $f^a/m$ orthogonal to $u^a$. (This
projection will actually be unnecessary here in the electromagnetic
case when we obtain the point particle equations of motion, since our
final answer for $f^a$ as defined by eq.~(\ref{deff}) will turn out to
be orthogonal to $u^a$, corresponding to $m$ being independent of
$\tau$ in eq.~(\ref{pmv}).) We have
\begin{eqnarray}
f^a &=& \frac{d}{d\tau} \left(\int_{\Sigma(\tau)} (T_{\rm body}^{ab}) 
                     \epsilon_{bcde} \right)
\nonumber \\ 
 &=& \int_{\Sigma(\tau)} \pounds_w \left( (T_{\rm body}^{ab}) 
                     \epsilon_{bcde}\right),
\label{flie}
\end{eqnarray}
where~$w^a$ is the vector field which generates the map between the
successive spatial slices~$\Sigma(\tau)$. Note that the value of the
integral does not depend upon how we choose to identify successive
spatial slices, i.e., it depends only on the normal component of $w^a$
(the lapse function) and not on its spatial projection into
$\Sigma(\tau)$ (the shift vector). Using the standard identity
\begin{equation}
\pounds_w \mu = w \cdot d\mu + d(w \cdot \mu)
\label{formsid}
\end{equation}
on a differential form $\mu$ (where the centered dot denotes
contraction into the first index of the form) together with Stokes'
theorem, we can rewrite the integrand to obtain
\begin{equation}
f^a =  \int_{\Sigma(\tau)}  \nabla_b T_{\rm body}^{ab} w^c \,d\Sigma_c
\label{eq:lieresult}
\end{equation}
where we now have rewritten our volume integral in more standard
notation by writing $w^c d\Sigma_c$ in place of $w^c \epsilon_{cdef}$.

We now assume that the body is coupled to a classical electromagnetic
field, $F_{ab}$, which satisfies Maxwell's equations with source given
by the charge-current density, $J^a$, of the body. Thus, $F_{ab}$
includes both the ``self-field'' of the body and any ``background''
electromagnetic field which may be present.  We denote the
stress-tensor of the Maxwell field by $T^{ab}_{\rm{EM}}$.  For further
generality, we also shall allow the body to be coupled to additional
classical matter (which does not couple to the electromagnetic field)
with stress-energy tensor $T^{ab}_{\rm{ext}}$. (This additional matter
can be thought of as a ``hand'' or ``string'' which pulls on the
charged body. We will set $T^{ab}_{\rm{ext}} = 0$ when we wish to
obtain the equations of motion for a charged body acted upon only by
an electromagnetic field, but permitting this extra coupling will
allow us to also obtain an expression for the radiation reaction force
for arbitrary world lines, when the charge is being ``pulled''.) By
conservation of total stress-energy, we have
\begin{equation}
\nabla_b [T_{\rm body}^{ab} + T^{ab}_{\rm{EM}} + T^{ab}_{\rm{ext}}] = 0.
\end{equation}
By Maxwell's equations, we have
\begin{equation}
\nabla_b T^{ab}_{\rm{EM}} = - F^{ab} J_b ,
\end{equation}
and therefore we can rewrite eq.~(\ref{eq:lieresult}) as
\begin{equation}
f^a = \int_{\Sigma(\tau)}  F^{ac} J_c w^b \,d\Sigma_b + f_{\rm ext}^a
\label{newf}
\end{equation} 
where
\begin{equation}
f_{\rm ext}^a \equiv  
	- \int_{\Sigma(\tau)}  \nabla_b T^{ab}_{\rm{ext}} w^c \,d\Sigma_c
\end{equation} 
For a body which accelerates with four-acceleration $a^b$, the lapse
function is given by
\begin{equation}
N \equiv w^a u_a = 1+r^\gamma a_\gamma
\end{equation}
where here we have extended the definition of $u^a$ over the
hypersurface by parallel transport (i.e., $u^a$ is the unit normal to
$\Sigma(\tau)$) and $r^\gamma$ are the Cartesian coordinate components
of the displacement vector on $\Sigma(\tau)$ with origin at the
representative world line. Thus, we obtain
\begin{equation}
f_{\rm EM}^\alpha \equiv f^\alpha - {f^\alpha}_{\rm{ext}}
	= \int_{\Sigma(\tau)}  F^{\alpha\beta} J_\beta N\,dV 
\end{equation}
where $dV$ denotes the ordinary volume element in the Euclidean
3-space $\Sigma(\tau)$ and the Greek indices denote components in a
global inertial coordinate system.

Let us now simplify the situation considerably further by assuming
that the body is exactly spherically symmetric and ``at rest'' with
respect to $u^a$ at ``time'' $\Sigma(\tau)$, so that $J^a = \rho(r)
u^a$ on $\Sigma(\tau)$. (However, we make no symmetry or other
assumptions concerning the electromagnetic field.)  Then we have
\begin{equation}
f_{\rm EM}^\alpha  = \int_{\Sigma(\tau)} \rho(r) E^\alpha N\,dV,
\label{fems}
\end{equation}
where~$E^\alpha \equiv F^{\alpha\beta} u_\beta$. We define~$q(r)$ to
be the total charge within radius~$r$ so that
\begin{equation}
\hat{r}^\beta D_\beta q(r) = 4\pi r^2 \rho(r),
\end{equation}
Then, we have
\begin{eqnarray}
f_{\rm EM}^\alpha
&=& \int_{\Sigma(\tau)} \frac{\hat{r}^\beta D_\beta q(r)}{4\pi r^2}
	E^\alpha N\,dV
\nonumber \\
&=& \int_{\Sigma(\tau)} D_\beta \left(\frac{\hat{r}^\beta q(r)}{4\pi r^2}
	E^\alpha N\right)\,dV
	- \int_{\Sigma(\tau)} \frac{q(r)}{4\pi r^2}
	\hat{r}^\beta D_\beta (E^\alpha N)\,dV
\nonumber \\ &&
	- \int_{\Sigma(\tau)} D_\beta \left(\frac{\hat{r}^\beta}{4\pi r^2}
	\right) q(r) E^\alpha N\,dV
\nonumber \\
&=& \int_{r=R} \frac{q(R)}{4\pi R^2} E^\alpha N\,dS
	- \int_{\Sigma(\tau)} \frac{q(r)}{4\pi r^2} 
	\left[\frac{\partial E^\alpha}{\partial r} N + 
	\hat{r}^\beta a_\beta E^\alpha\right] \,dV
\nonumber \\
&&
	- \int_{\Sigma(\tau)} \delta^3(r) q(r) E^\alpha N\,dV \nonumber \\
&=& e \langle E^\alpha N\rangle_R
	- \int_{\Sigma(\tau)} \frac{q(r)}{4\pi r^2} 
	\left[\frac{\partial E^\alpha}{\partial r} N + 
	\hat{r}^\beta a_\beta E^\alpha\right] \,dV,
\label{fEM}
\end{eqnarray}
where~$R$ represents the outer radius of the charged body,~$e \equiv
q(R)$ is the total charge, and~$\langle \rangle_R$ denotes the average
over the sphere~$r=R$.

The difficulties encountered in trying to obtain a prescription
for~$f_{\rm EM}^a$ in a simple, straightforward manner by taking the
point particle limit of the equations of motion for an extended body
can be seen directly from eq.~(\ref{fEM}).  Let us split the total
electromagnetic field, $F_{ab}$, into a smooth ``background piece'',
$F^{(0)}_{ab}$, which has a smooth limit as the size of the body is
shrunk to zero and a piece $F^{(1)}_{ab}$ which we may view as being
``due to the charge'' itself. (Most commonly $F^{(1)}_{ab}$ would be
taken to be the retarded solution with source $J^a$.) When the size of
the body becomes sufficiently small (in particular, much smaller than
the scale of variation of $F^{(0)}_{ab}$), the contribution of
$F^{(0)}_{ab}$ to the volume integral in eq.~(\ref{fEM}) becomes
negligible, whereas the surface term straightforwardly yields the
Lorentz force law
\begin{equation}
f_{\rm EM}^{(0)a} = e F^{(0)ab} u_b.
\end{equation}
since the lapse function, $N$, may be approximated as $1$ when the
body shrinks to zero size with $E^a$ remaining bounded.  Indeed, even
if the body were nonspherical and/or not perfectly ``at rest'', there
would be no significant difficultly in obtaining the lowest order
finite size corrections to $f_{\rm EM}^{(0)a}$ (see, e.g., Dixon
\cite{dixon64}).

However, the situation is completely different with regard to
$F^{(1)}_{ab}$, which contributes a ``self-force'' $f_{\rm EM}^{(1)a}$
of order $e^2$. If the total charge is fixed, $F^{(1)}_{ab}$ and its
spatial derivatives become unboundedly large as the size of the body
is made small, so that the integrals appearing in eq.~(\ref{fEM})
cannot easily be controlled. These integrals could still have a well
defined limit as a result of cancellations over the different portions
of the body, but the situation clearly is extremely delicate. In
particular, small deviations from exact sphericity or being exactly
``at rest'' (as well as small corrections due to curvature when we
consider the motion of charged bodies in curved spacetime) could
easily contribute finite corrections to $f_{\rm EM}^{(1)a}$. Indeed,
corrections to eq.~(\ref{pmv}) itself could be large in the point
particle limit.

The singular behavior of the self-field in the point particle limit
makes it very difficult to directly extract from (\ref{fEM}) any
guidance as to what the electromagnetic force should be, even in the
regime where one would expect this force to be essentially composition
independent. However, inspection of eq.~(\ref{fEM}) provides a
possible means of dealing with this difficulty, and this comprises
the key new idea of this paper: Suppose that, rather than calculating
the electromagnetic force (\ref{fEM}) on a particular body, we instead
attempt to calculate the {\em difference} in the electromagnetic force
on two bodies of the same (or very similar) composition, which move on
different world lines in (possibly different) spacetimes. Then, under
appropriate circumstances (see below), it seems plausible that we may
identify neighborhoods of the bodies in such a way that the {\em
difference} in the electromagnetic field and its relevant first
spatial derivatives for the two bodies can remain bounded as both of
the bodies shrink to zero size at the same rate. If so, the volume
integral contribution to the {\em difference} in $f_{\rm EM}^a$ for
the two bodies (the second term on the right side of eq.~(\ref{fEM}))
will go to zero. The difference in $f_{\rm EM}^a$ for the two bodies
will then be given by a version of the Lorentz force law, wherein we
take the difference in the electromagnetic fields, average this
difference over the surface of the (identified) bodies as in the first
term on the right side of eq.~(\ref{fEM}) (with $N = 1$), and then let
the bodies shrink to zero size.  Thus, if we consider the difference
in electromagnetic forces between two bodies rather than the force on
a single body, the ``point particle limit'' should be much less
delicate, and, in particular, much less sensitive to small deviations
from sphericity, etc. (provided, of course, that these deviations are
essentially the same for both bodies).

Indeed, even if the bodies are in different curved spacetimes, it
should be possible to keep the difference in the electromagnetic
forces on these (suitably identified) bodies under good control. In a
curved spacetime, the above calculations would be modified in the
following ways. First, we must use parallel transport to define the
integral expression for $p^a$. To do this explicitly, it is convenient
to introduce an arbitrary unit vector, $k^a$, at a point on the
representative world line and then parallel transport $k^a$ along the
worldline. For each $\tau$, let $\Sigma(\tau)$ be the hypersurface
generated by geodesics orthogonal to $u^a$ at point $z(\tau)$ of the
representative world line. For each $\tau$, we define $k^a$ on
$\Sigma(\tau)$ by parallel transport along these geodesics, thereby
defining $k^a$ in a neighborhood of the entire worldline.  We define
the four-momentum $p^a$ at point $z(\tau)$ by
\begin{eqnarray}
k_a p^a &=& \int_{\Sigma} k_a (T_{\rm body}^{ab}) \epsilon_{bcde},
\label{defpcs}
\end{eqnarray}
Again, we assume that a representative world line can be found for the
extended body such that eq.~(\ref{pmv}) holds to an excellent
approximation when the body is sufficiently small and nearly
spherical.

If we introduce Riemann normal coordinates at point $z(\tau)$ in place
of the Cartesian coordinates of the flat spacetime derivation, the
calculation which produced eq.~(\ref{fEM}) from eq.~(\ref{flie}) is
essentially unchanged, with the curvature of the background
introducing only a few corrections to eq.~(\ref{fEM}). First, a new
term involving $\nabla_b k_a$ explicitly appears in the integrand in
eq.~(\ref{eq:lieresult}). Further corrections also result from the
fact that the Riemann normal coordinate components of $k^a$ are not
constant. In addition, $w^b d\Sigma_b$ deviates slightly from $N dV$,
where $dV$ is the Riemann normal coordinate volume element for the
hypersurface.  However, these corrections all decrease with the size
of the particle and should become negligible in the point particle
limit $R \rightarrow 0$.

When will two bodies (possibly in different curved spacetimes) be such
that the difference in their electromagnetic fields will be suitably
bounded as their size shrinks to zero? To answer this question
properly, we would need to carefully examine the behavior of the
``self-field'' of extended bodies as the ``point particle limit'' is
approached. We shall not attempt to analyze this here. However, a good
guess as to the answer to this question can be obtained by examining
the exterior field of a point charge in a curved spacetime, and
finding the conditions under which the difference between the fields
of two such charges is---with a suitable identification of
neighborhoods of the world lines of the particles---suitably bounded
as one approaches a point on the world line of the particle.

To do so, we need to study the singular behavior of the
electromagnetic field of a point charge in curved spacetime as one
approaches the world line of the point charge. As stated above, we are
concerned only with solutions to eq.~(\ref{Ftotal}) which are singular
precisely on the world line of the particle itself. It follows from
the general theory of propagation of singularities (see theorem 26.1.1
of Hormander \cite{hormander}), that all such solutions have the same
singular behavior, i.e., the difference between any two solutions must
be smooth on the world line of the particle itself.\footnote{Note that
this implies, in particular, that the advanced minus retarded solution
is always smooth throughout the spacetime---including on the world
line of the particle---provided only that the advanced and retarded
solutions themselves are nonsingular off of the worldline of the
particle. (However, as illustrated by the example of a uniformly
accelerating charge in Minkowski spacetime, the advanced and retarded
solutions need not always be nonsingular away from the worldline of
the particle.)} Thus, to examine the singular behavior, it suffices to
focus attention on any particular solution. When $(M, g_{ab})$ is
globally hyperbolic, it is convenient to examine the advanced and
retarded solutions. The behavior of these solutions near the world
line of the particle can be calculated by the Hadamard expansion
techniques detailed by Dewitt and Brehme \cite{db}. The result is
\begin{eqnarray}
F^\pm_{a' b'}(x) = &2e& 
\bar{g}_{a' [a}\bar{g}_{|b'|  b]} 
[  r^{-2} \kappa^{-1} u^{a}  \Omega^{b}
 + \frac{1}{2} r^{-1} \kappa^{-3} a^{a} u^{b}
 + \frac{1}{8} \kappa^{-5} u^{a}  \Omega^{b}   a^2
\nonumber\\
{}&-&\frac{1}{2} \kappa^{-3} \dot{a}^{a} \Omega^{b}
\pm \frac{2}{3} \kappa^{-4} \dot{a}^{a} u^{b}
 +  \frac{1}{12} \kappa^{-1} u^{a}  \Omega^{b} R
 -  \frac{1}{6} \kappa^{-1} u^{a} R^{b}{}_{c} \Omega^{c}
\nonumber\\
{}&+&\frac{1}{2} \kappa^{-1} \Omega^{a} R^{b}{}_{c} u^{c}
 +  \frac{1}{12} \kappa^{-1} u^{a} \Omega^{b} 
                 R_{cd} \Omega^{c} \Omega^{d}
 +  \frac{1}{2}  \kappa^{-1} R^{a}{}_{c}{}^{b}{}_{d}
                             u^{c} \Omega^{d}
\nonumber \\
{}&-& \frac{1}{12} \kappa^{-3} u^{a} \Omega^{b} R_{cd}
                               u^{c} u^{d}
 +  \frac{1}{6} \kappa^{-3} u^{a} R^{b}{}_{cde}
                            u^{c} u^{d} \Omega^{e}
\nonumber\\
{}&\mp&\frac{1}{3} \kappa^{-2} u^{a} R^{b}{}_{c} u^{c}]
\nonumber \\
&\pm& e \int^{\pm\infty}_{\tau^\pm} 2 \nabla_{[b'}  G^\pm_{a']a''}
                u^{a''}(\tau'')\,d\tau''
 + O(r), 
\label{dbeqn}
\end{eqnarray}
Here, $u^a$ denotes the four-velocity of the point charge at point
$z(\tau)$ on its world line, and $x$ denotes a point sufficiently near
$z(\tau)$ lying on the hypersurface generated by geodesics from
$z(\tau)$ which are orthogonal to $u^a$. The outward-directed unit
tangent at $z(\tau)$ to the geodesic passing through $x$ is denoted by
$\Omega^a$, and the affine parameter of $x$ (i.e., the distance of $x$
from the world line) is denoted by $r$. Primed indices refer to
tensors at $x$, while unprimed indices refer to tensors at $z(\tau)$,
and $\bar{g}^{a'}{}_a(x,z(\tau))$ denotes the bitensor of geodesic
parallel transport.  The quantity $\kappa$ is defined by $\kappa
\equiv \sqrt{- u^a u^b \nabla_a \nabla_b \sigma - a^a \nabla_a
\sigma}$, where~$\sigma$ denotes the {\em biscalar of squared geodesic
distance\/}, which plays a fundamental role in all of these
expansions. (The normalization of $\sigma$ is such that
$\sigma(x,z(\tau)) = r^2/2$.) We have set $a^a \equiv u^b \nabla_b
u^a$ and $\dot{a}^{a} \equiv u^b \nabla_b a^a$. The last term in the
above equation is usually referred to as the ``tail term'', and it
results from the failure of Huygen's principle in curved spacetime. In
that term, $G^\pm_{a'a''}(x,z(\tau''))$ denotes the advanced/retarded
Green's function for the vector potential in the Lorentz gauge, so
that
\begin{equation}
\nabla^b \nabla_b G^\pm_{aa'} - R_a{}^b G^\pm_{ba'} 
	= - 4 \pi \bar{g}_{aa'} \delta(x,z),
\end{equation}
and $\tau^\pm$ denotes the proper time of the point on the world line
of the charged particle which intersects the future/past light cone of
$x$. (The alternating sign in front of the tail term integral in
eq.~(\ref{dbeqn}) merely puts the limits in the appropriate time
order.) The integral in the tail term runs from $(\tau^\pm \pm
\epsilon)$ to $\pm \infty$ with the limit $\epsilon \rightarrow 0$
then being taken.\footnote{If the worldline is not complete (i.e., if
it does not extend to infinite proper time in the future/past), then
the upper limit of the tail term integral should be the
maximum/minimum proper time values of the curve.}  No distributional
component of $G^\pm_{a'a''}(x,z(\tau''))$ is encountered in the
integral, and we shall assume that the tail term remains smooth as $x$
approaches $z(\tau)$ (as should be the case if suitable asymptotic
conditions are placed on the world line and the spacetime).

In addition to obvious notational differences, our expression differs
from eq.~(5.12) of Dewitt and Brehme \cite{db} in three ways. First,
we have added the terms that Dewitt and Brehme omitted due to a
trivial calculational error (see Hobbs \cite{hobbs}). Second, we have
written the tail term in terms of the Green's function
$G^\pm_{a'a''}(x,z(\tau''))$ rather than the Hadamard expansion term
$v_{a'a''}(x,z(\tau''))$. The latter expression gives the correct form
of the tail term only when $x$ is sufficiently close to $z(\tau'')$,
and, in general, is not even defined for large separations (when, in
particular, there need not be a unique geodesic joining~$x$
and~$z(\tau'')$). Finally, our sign convention for the Riemann tensor
is that of \cite{wald}, which is opposite to that of Dewitt and
Brehme.

In order to find the singular behavior of $F^\pm_{a'b'}$, we need to
expand the coordinate components of $\bar{g}_{a'a}$ and~$\kappa$. Both
of these quantities take particularly simple forms in terms of Riemann
normal coordinates based at $z(\tau)$. We have
\begin{equation}
\bar{g}^{\alpha'}{}_\alpha = \delta^{\alpha'}{}_\alpha + 
	\frac{1}{6} r^2 \Omega^\gamma \Omega^\delta
	R^{\alpha'}{}_{\gamma\alpha\delta} + O(r^3)
\end{equation}
and
\begin{equation}
\kappa = \sqrt{1 + r a^\alpha \Omega_\alpha 
	+\frac{1}{3} r^2 u^\alpha u^\beta \Omega^\gamma \Omega^\delta
	R_{\alpha\gamma\beta\delta} + O(r^3)}.
\end{equation}
Substituting these expansions into eq.~(\ref{dbeqn}), we have
\begin{eqnarray}
F^\pm_{{\alpha'}{\beta'}} (x)
&=& 2e[ r^{-2} u_{[\alpha'} \Omega_{\beta']} 
       -\frac{1}{2}r^{-1} (a^\alpha \Omega_\alpha) 
                          u_{[\alpha'} \Omega_{\beta']} 
       +\frac{3}{8}(a^\alpha \Omega_\alpha)^2 u_{[\alpha'} \Omega_{\beta']} 
       +\frac{1}{2} r^{-1} a_{[\alpha'} u_{\beta']}
\nonumber\\ &&\phantom{e[}
       -\frac{1}{6} u^\alpha u^\beta \Omega^\gamma \Omega^\delta
                    R_{\alpha\gamma\beta\delta} u_{[\alpha'} \Omega_{\beta']}
       -\frac{3}{4} (a^\alpha\Omega_\alpha) a_{[\alpha'} u_{\beta']}
       +\frac{1}{6} \Omega_{[\beta'} R_{\alpha']\sigma\alpha\tau}
                    u^\alpha \Omega^\sigma \Omega^\tau
\nonumber \\ &&\phantom{e[}
 +  \frac{1}{8}  u_{[\alpha'} \Omega_{\beta']}   a^2
 -  \frac{1}{2}  \dot{a}_{[\alpha'} \Omega_{\beta']}
\pm \frac{2}{3}  \dot{a}_{[\alpha'} u_{\beta']}
 +  \frac{1}{12} u_{[\alpha'}  \Omega_{\beta']} R
 -  \frac{1}{6}  u_{[\alpha'} R_{\beta']\gamma} \Omega^\gamma
\nonumber \\ &&\phantom{e[}
 +  \frac{1}{2}  \Omega_{[\alpha'} R_{\beta']\gamma} u^\gamma
 +  \frac{1}{12} u_{[\alpha'} \Omega_{\beta']} R_{\gamma\delta} 
                 \Omega^\gamma \Omega^\delta
 +  \frac{1}{2}  R_{[\alpha'|\gamma|\beta']\delta} u^\gamma \Omega^\delta
\nonumber \\ &&\phantom{e[}
 -  \frac{1}{12} u_{[\alpha'} \Omega_{\beta']} 
                 R_{\gamma\delta} u^\gamma u^\delta
 +  \frac{1}{6}  u_{[\alpha'} R_{\beta']\gamma\delta\epsilon} 
                 u^\gamma u^\delta \Omega^\epsilon
\mp \frac{1}{3}  u_{[\alpha'} R_{\beta']\gamma} u^\gamma]
\nonumber \\ &&\phantom{e[}
\pm e \int^{\pm\infty}_{\tau\pm} \nabla_{[\beta'}  G^\pm_{\alpha']\alpha''}
                u^{\alpha''}(\tau'') \, d\tau''
+ O(r)
\label{Fexpand}
\end{eqnarray}
Although this formula is explicitly for the advanced/retarded solution
in a globally hyperbolic spacetime, as noted above, the singular
behavior of $F_{ab}$ will be the same as in eq.~(\ref{Fexpand}) for
any solution of Maxwell's equations with source (\ref{Ftotal}) in a
(possibly non-globally-hyperbolic) spacetime, provided only that
$F_{ab}$ is smooth away from the world line of the particle.

From eq.~(\ref{Fexpand}), it can be seen that the divergent terms in
$F^\pm_{a'b'}$ depend only upon the four-velocity and the
four-acceleration of the world line at~$z(\tau)$. In particular, they
do not depend upon the spacetime curvature or derivatives of the
acceleration.  Furthermore, although many of the finite terms (which
do depend upon the curvature and $\dot{a}^a$) are direction-dependent
and thus have singular angular derivatives on the world line, the
radial derivatives of these terms (which is all that enters the volume
term in eq.~(\ref{fEM})) are bounded.  Therefore, it seems plausible
that if we have two bodies with the same magnitude of acceleration at
corresponding points $P$ and $\widetilde{P}$ on their representative
world lines and if we identify neighborhoods of $P$ and
$\widetilde{P}$ using Riemann normal coordinates, with $u^a$
aligned with $\tilde{u}^a$ and $a^a$ aligned with $\tilde{a}^a$, then
the singular contributions of the ``self-fields'' to $f_{\rm EM}^a$ in
the point particle limit should cancel. Thus, the difference in
$f_{\rm EM}^a$ for the two bodies in the point particle limit should
be given by a version of the Lorentz force law wherein we average the
difference in the electromagnetic fields over a surface of radius $r$
as in the first term on the right side of eq.~(\ref{fEM}) (with $N =
1$), and then let $r \rightarrow 0$. This provides the motivation for
Axiom~\ref{ax:comp} below.

\subsection{The axioms}
\label{sec:emax}

We now are ready to state our main axiom, the motivation for which was
given in the previous subsection.

\begin{emaxiom}[Comparison axiom] \label{ax:comp}
Consider two points,~$P$ and~$\widetilde{P}$, each lying on timelike
world lines in possibly different spacetimes which contain Maxwell
fields $F_{ab}$ and $\widetilde{F}_{ab}$ sourced by particles of
charge $e$ on the world lines. If the four-accelerations of the world
lines at~$P$ and~$\widetilde{P}$ have the same magnitude, and if we
identify the neighborhoods of~$P$ and~$\widetilde{P}$ via Riemann
normal coordinates such that the four-velocities and
four-accelerations are identifed, then the difference in the
electromagnetic forces~$f_{\rm EM}^a$ and~$\tilde{f}_{\rm EM}^a$ is
given by the limit as~$r \rightarrow 0$ of the Lorentz force
associated with the difference of the two fields averaged over a
sphere at geodesic distance~$r$ from the world line at $P$.
\begin{equation}
f_{\rm EM}^a - \tilde{f}_{\rm EM}^a = \lim_{r \rightarrow 0} \left(
	\langle F^{ab} - \widetilde{F}^{ab} \rangle
	\right) u_b
\end{equation}
\end{emaxiom}

Axiom~\ref{ax:comp} is a very powerful one, since it enables us to
compute the difference in electromagnetic force between any two
particles which have the same instantaneous acceleration. Thus, to
obtain $f_{\rm EM}^a$ for an arbitrary trajectory in an arbitrary
curved spacetime, it suffices to know $f_{\rm EM}^a$ for a uniformly
accelerating particle---with arbitrary acceleration $a^a$---in
Minkowski spacetime, with the electromagnetic field chosen to be, say,
the half-advanced, half-retarded solution.

Let us, then, consider this special case. By symmetry, $f_{\rm EM}^a$
must be proportional to $a^a$. If the proportionality factor were
constant, such a force would correspond merely to a ``mass
renormalization'', and could be redefined away. On the other hand,
such a redefinition would not be possible if the proportionality
factor varied with acceleration. We see no argument from symmetry
considerations alone which would forbid the presence of such a term.
However, this spacetime, world line, and Maxwell field possess a time
reversal symmetry about each point on the world line, which suggests
that the particle always should be absorbing as much electromagnetic
energy as it radiates, so the electromagnetic field should be doing
``no net work'' on the particle.  This, in turn, strongly suggests
that $f_{\rm EM}^a = 0$ in this case. Indeed, if we did not have
$f_{\rm EM}^a = 0$, the type of calculation given in section 17.2 of
Jackson \cite{jackson} would show that our resulting prescription for
$f_{\rm EM}^a$ would fail to conserve energy for a point particle
trajectory which begins and ends in inertial motion (where, in this
calculation, the infinite self-energy of the Coulomb field of the
particle is discarded at the initial and final times).  This motivates
the following additional axiom, which agrees with standard claims made
in textbooks (see, e.g., Jackson \cite{jackson}):

\begin{emaxiom}[Flat spacetime axiom] \label{ax:flat}
If~$(M,g_{ab})$ is Minkowski spacetime, the world line is uniformly
accelerating, and~$F_{ab}$ is the half-advanced, half-retarded
solution, $F_{ab} =\frac{1}{2}[F^-_{ab} + F^+_{ab}]$, then~$f^a=0$ at
every point on the world line.
\end{emaxiom} 

Note that, since the advanced and retarded solutions for $F_{ab}$ for
a uniformly accelerating charge in Minkowski spacetime coincide in a
neighborhood of the worldline (indeed, within the entire ``Rindler
wedge'' containing the worldline \cite{boulware}), it follows
immediately from Axiom~\ref{ax:comp} that we also have $f^a = 0$ when
$F_{ab}$ is given by the advanced solution, $F^+_{ab}$, or by the
retarded solution, $F^-_{ab}$. Thus, we would obtain an equivalent
axiom if we replaced the half-advanced, half-retarded solution by the
advanced solution or the retarded solution.

In the next section, we shall use Axioms~\ref{ax:comp}
and~\ref{ax:flat} together with eq.~(\ref{Fexpand}) to compute $f_{\rm
EM}^a$ for an arbitrary charged particle trajectory in an arbitrary
curved spacetime.

\subsection{The prescription}
\label{sec:empres}

Let $P$ be a point on the world line of a charged particle in a curved
spacetime $(M, g_{ab})$ containing a Maxwell field $F_{ab}$ satisfying
eq.~(\ref{Ftotal}), where $F_{ab}$ is singular only on the world line
of the particle. For simplicity we assume that $(M, g_{ab})$ is
globally hyperbolic so a unique retarded Green's function exists; as
explained at the end of this subsection, our formulas can easily be
generalized to the non-globally-hyperbolic case.  Let $u^a$ denote the
four-velocity of the world line at $P$ and let $a^a$ denote its
acceleration at $P$.

We may view the tangent space at $P$ as a copy of Minkowski spacetime.
We shall denote the origin of this tangent space by
$\widetilde{P}$. In this Minkowski spacetime, consider a uniformly
accelerating trajectory passing through $\widetilde{P}$ with
four-velocity $\tilde{u}^a = u^a$ and acceleration $\tilde{a}^a =
a^a$. By Axiom~\ref{ax:flat}, the electromagnetic force on this
uniformly accelerating Minkowski trajectory vanishes when the
electromagnetic field is given by the (Minkowski) half-advanced,
half-retarded solution.

In $(M, g_{ab})$, we write $F^{\rm in}_{ab} = F_{ab} - F^-_{ab}$,
where $F^-_{ab}$ denotes the retarded solution, and it is assumed that
$F^{\rm in}_{ab}$ is smooth on the world line of the particle. Near
the actual trajectory of the particle (in $(M, g_{ab})$), $F^-_{ab}$
is given by eq.~(\ref{Fexpand}). On the other hand, near the uniformly
accelerating trajectory in the tangent space at $\widetilde{P}$, the
(Minkowski) half-advanced, half-retarded solution $\widetilde{F}_{ab}
\equiv \frac{1}{2}({\widetilde{F}^-}_{ab} + {\widetilde{F}^+}_{ab})$
also is given by eq.~(\ref{Fexpand}) except that the ``tail term'' and
all of the terms involving the curvature are absent, and there is
cancellation of terms involving $\dot{a}$ and
$a^2$. Axiom~\ref{ax:comp} instructs us to subtract this Minkowski
retarded solution from $F_{ab}$ (using the exponential map---or,
equivalently, Riemann normal coordinates---to compare them), average
this difference over a sphere of radius $r$, and then let $r
\rightarrow 0$. The electromagnetic force on the particle at $P$ is
then just the Lorentz force associated with the resulting field. We
obtain
\begin{eqnarray}
f_{\rm EM}^a &=& e(F^{\rm in})^{ab}u_b + \frac{2}{3} e^2 (\dot{a}^a - a^2 u^a) 
	+\frac{1}{3} e^2(R^a{}_b u^b + u^a R_{bc}u^b u^c) \nonumber \\
&& + e^2u_b\int_{-\infty}^{\tau^-}
	\nabla^{[b} (G^-)^{a]c'} u_{c'}(\tau')\,d\tau'
\label{eqmotion}
\end{eqnarray}
The corresponding equation of motion of a charged particle subject to
no additional (i.e., non-electromagnetic) external forces is then
simply $f_{\rm EM}^a = ma^a$.

Our result (\ref{eqmotion}) agrees with that of Dewitt and
Brehme~\cite{db} as corrected by Hobbs~\cite{hobbs}. Although we, of
course, made crucial use of the Hadamard expansion for the retarded
Green's function (\ref{Fexpand}), no other lengthy computations were
needed in our approach, since we did not need to compute the behavior
of the electromagnetic stress-energy tensor near the world line of the
particle.

Note that the first term in (\ref{eqmotion}) is the ordinary Lorentz
force due to the incoming field. The second term corresponds to the
familar flat spacetime Abraham-Lorentz damping term. The third term is
a local curvature term, whose presence is necessary to maintain
conformal invariance of $f_{\rm EM}^a$. Finally, the fourth term is
the so-called ``tail term'' resulting from the failure of Huygen's
principle in curved spacetime.

Due to the presence of the Abraham-Lorentz term, the equation of
motion $f_{\rm EM}^a = ma^a$ shares the unphysical ``runaway
solutions'' of the ordinary flat spacetime equation of motion. As in
the flat spacetime case, this difficulty can be resolved through the
reduction of order technique. An exposition of the rational for this
technique as well as an explanation of how to implement it in a
general context can be found in section IV~D of \cite{flanwald}. To
implement it here, we view $\epsilon \equiv e^2/m$ as a ``small
parameter''. We differentiate eq.~(\ref{eqmotion}) (with $f_{\rm
EM}^a$ set equal to $ma^a$) to obtain an expression for $\dot{a}^a$,
and then substitute this expression back in eq.~(\ref{eqmotion}),
neglecting terms which are higher than first order in $\epsilon$. We
then similarly eliminate the terms involving $a^a$ from the right side
of eq.~(\ref{eqmotion}). The result is
\begin{eqnarray}
a^a &=& \frac{e}{m} (F^{\rm in})^{ab} u_b 
      + \frac{2}{3} \frac{e^2}{m} \left(\frac{e}{m} u^c \nabla_c
      (F^{\rm in})^{ab} u_b + \frac{e^2}{m^2} (F^{\rm in})^{ab} F^{\rm in}_{bc}
      u^c - \frac{e^2}{m^2} u^a (F^{\rm in})^{bc} u_c F^{\rm in}_{bd} u^d
      \right) \nonumber \\
&& +\frac{1}{3} \frac{e^2}{m}(R^a{}_b u^b + u^a R_{bc}u^b u^c)
	+ \frac{e^2}{m} u_b \int_{-\infty}^{\tau^-} 
	\nabla^{[b} (G^-)^{a]c} u_{c}(\tau')\,d\tau'
\label{reduct}
\end{eqnarray}
We believe that this equation properly describes the motion of a
small, nearly spherical charged body in a curved spacetime, taking
into account the leading order effects of the body's ``self-field''.

Inasmuch as they require the retarded solution to be singled out,
expressions (\ref{eqmotion}) and (\ref{reduct}) are applicable as they
stand only for a particle in a globally hyperbolic spacetime. However,
since Axiom~\ref{ax:comp} did not require global hyperbolicity, it is
clear that our axioms also determine the electromagnetic force and
equations of motion in the non-globally-hyperbolic case as
well. Perhaps the simplest way of generalizing our formulas to the
non-globally-hyperbolic case is as follows: If we wish to obtain
$f_{\rm EM}^a$ at a point $z(\tau)$ on the world line of a charged
particle in a non-globally-hyperbolic spacetime, simply choose a
(sufficiently small) globally hyperbolic neighborhood of
$z(\tau)$. Eqs. (\ref{eqmotion}) and (\ref{reduct}) then hold at
$z(\tau)$, where $F^{\rm in}_{ab}$ and the tail term are defined in
the appropriate manner, relative to that neighborhood.

\section{Gravitational radiation reaction}
\label{sec:gr}

In this section, we seek to obtain the gravitational analog of our
formula (\ref{eqmotion}) above for the total electromagnetic force
(including radiation reaction) on a charged particle, as well as the
analog of our equation of motion (\ref{reduct}) above. The latter will
provide us with the lowest order correction to geodesic motion of a
particle resulting from radiation reaction effects. In our approach,
we shall not make any of the slow motion or post-Newtonian
approximations common to most other treatments of gravitational
radiation reaction. On the other hand, the applicability of our
results will be limited to the motion of a small, nearly spherical
body.

There are many physical and mathematical similarities in the analyses
of the electromagnetic and gravitational radiation reaction forces,
and our analysis of gravitational radiation reaction will ultimately
closely parallel that of the electromagnetic case. However, there also
are a number of very significant differences between these two
cases. We begin our analysis of the gravitational case by explaining
in detail the nature of these differences.

Probably the most significant difference between the electromagnetic
and gravitational cases concerns the formulation of the question which
we would like to pose.  As discussed in detail in the Introduction, we
do not view a ``point particle'' as a fundamental object, but,
instead, view the ``point particle limit'' as a convenient
mathematical means of summarizing results concerning the behavior of
one-parameter families of extended body solutions in the limit where
not only the size but also the charge and mass of the body go to zero
in a suitable manner. Nevertheless, in the electromagnetic case, there
is no difficulty in making sense of solutions to Maxwell's equations
if we let the size of the body shrink to zero keeping its charge
fixed. This enabled us to pose (and propose an answer to) the
following idealized question in Sec~\ref{sec:em}: Given a solution to
eq.~(\ref{Ftotal}) for a Maxwell field with point particle source,
what is the total electromagnetic force on the charged particle?  On
the other hand, the corresponding question in the gravitational case
would be: Given a solution to Einstein's equation with a point
particle source, what is the total ``gravitational force'' on the
particle?  However, as already noted in the Introduction, this
question makes no sense, since there is no notion of a solution to
Einstein's equation with a ``point mass'' source \cite{gt}.

A resolution of this difficulty is suggested by the fact that we are
really interested in the case of (small) extended bodies whose
self-gravity is ``weak''.\footnote{However, we do not wish to preclude
the possibility of eventually extending our analysis to small bodies
with strong self-gravity; see \cite{hartle}.}Thus, it should be
adequate to treat the gravitational effects of the body via linearized
perturbation theory off of a background vacuum spacetime. For linear
equations, there is no difficulty in making sense of solutions with
distributional sources, so, when working with the linearized
equations, it becomes mathematically legal to let the size of the body
shrink to zero, keeping its mass fixed. This suggests that we pose the
following question, which is directly analogous to the question posed
in Sec~\ref{sec:em}: Let $(M, g^{(0)}_{ab})$ be a spacetime satisfying
the vacuum Einstein equation, let $z(\tau)$ be an arbitrary timelike
worldline in $(M, g^{(0)}_{ab})$, and let $\gamma_{ab}$ be a solution
of the linearized Einstein equation sourced by a particle following
this worldline. What is the total ``gravitational force'' on the
particle?

Unfortunately, the above question also suffers from serious
mathematical inconsistencies: By the linearized Bianchi identity, the
linearized Einstein equation implies exact conservation of the
stress-energy of the (linearized) source with respect to the
background metric. In the limit where the source is a point particle,
this conservation requires the world line of the particle to be a
geodesic of the background metric. Thus, if $z(\tau)$ is not chosen to
be a geodesic of $g^{(0)}_{ab}$, the above question makes no sense
since there does not exist any solution $\gamma_{ab}$ whatsoever to
the linearized Einstein equation with this source. But, a knowledge of
the total ``gravitational force'' only for geodesics of $g^{(0)}_{ab}$
would not be adequate for obtaining the self-consistent motion of the
particle under the influence of its own gravitational ``self-force'',
since such a particle will deviate from geodesic motion.

The origin of this difficulty can be understood as follows. Even for
an extended body with very weak self-gravity, the linearized Einstein
equation does not hold exactly; rather there are nonlinear corrections
to this equation. Although these nonlinear terms make only a very
small correction to $\gamma_{ab}$, it is precisely the presence of
these terms which are responsible for the deviations from geodesic
motion. By throwing away the nonlinear terms in $\gamma_{ab}$, we
exclude from the outset the possibility that the particle fails to
move a geodesic, thereby making it mathematically inconsistent to
study departures from geodesic motion.

To see this more explicitly, consider the exact Einstein equation for
the metric $g^{(0)}_{ab} + \gamma_{ab}$, written in the form of the
linearized Einstein equation for $\gamma_{ab}$ in the Lorentz gauge,
with the nonlinear terms in $\gamma_{ab}$ moved to the right side of
the equation (in a schematic manner) to aid us in viewing them as an
additional ``source term'':
\begin{eqnarray}
\nabla^{(0)c} \nabla^{(0)}_c \bar{\gamma}_{ab} 
	- 2 R^{(0)c}{}_{ab}{}^d \bar{\gamma}_{cd} 
	&=& -16 \pi T_{ab} + 
	[\mbox{nonlinear terms in $\gamma_{ab}$}] \label{exact} \\
\nabla^{(0)a} \bar{\gamma}_{ab} &=& 0 \label{lorentz}
\end{eqnarray}
where $\bar{\gamma}_{ab} \equiv \gamma_{ab} - \frac{1}{2} \gamma
g^{(0)}_{ab}$.

As already noted above, for a body with weak self-gravity, the matter
stress-energy $T_{ab}$ should dominate the ``nonlinear terms in
$\gamma_{ab}$''. More precisely, $T_{ab}$ is of order $m$, whereas if
there is no incoming gravitational radiation, the nonlinear terms
should have magnitude of order $m^2$ and higher, where $m$ denotes the
mass of the body. As we shall see in more detail below, a knowledge of
the resulting dominant $O(m)$ contribution to $\gamma_{ab}$ from
$T_{ab}$ will suffice for determining the leading order contribution
to the self-force, so we should make little error by dropping the
nonlinear terms. However, if we do so, there are no solutions to
eqs.(\ref{exact}) and (\ref{lorentz}) unless $\nabla^{(0)}_a T^{ab} =
0$.

However, a means of dealing with this difficulty is suggested by the
form in which we have written the equations. Even when $\nabla^{(0)}_a
T^{ab} \neq 0$, no mathematical inconsistencies occur in
eq.~(\ref{exact}) alone when the nonlinear terms are dropped. It is
only when the Lorentz gauge condition (\ref{lorentz}) is adjoined to
this equation that inconsistencies arise. Thus, we propose to simply
relax the Lorentz gauge condtion so that it holds only to the required
accuracy, i.e., to $O(m)$. (This can be ensured by simply requiring
that any ``incoming radiation'' contributions to $\gamma_{ab}$ satisfy
the Lorentz gauge condition.)  The resulting system of equations
should then have the accuracy needed to obtain the leading order
contribution to the gravitational self-force, but should not suffer
from the mathematical inconsistencies which would occur if the
linearized Einstein equation were used to relate $\gamma_{ab}$ to
$T_{ab}$. We note that our viewpoint appears to correspond to that
taken in \cite{mino}, and similar procedures for relaxing field
equations or gauge conditions at appropriate orders also occur in many
other approaches to obtaining self-consistent equations of motion
(see, e.g., \cite{eih}).

Having reformulated the equations for $\gamma_{ab}$ in this manner, we
now may consider the point particle limit and pose the following
question: Let $(M, g^{(0)}_{ab})$ be a spacetime satisfying the vacuum
Einstein equation, let $z(\tau)$ be an arbitrary timelike worldline in
$(M, g^{(0)}_{ab})$, and let $\gamma_{ab}$ be a solution of
\begin{equation}
\nabla^c \nabla_c \bar{\gamma}_{ab} - 2 R^c{}_{ab}{}^d \bar{\gamma}_{cd} 
	= - 16 \pi m \int \delta(x,z(\tau))\,u^b(\tau)\,d\tau,
\label{gravwave}
\end{equation}
where $\bar{\gamma}^{\rm in}_{ab} \equiv \bar{\gamma}_{ab} -
\bar{\gamma}^-_{ab}$ satisfies the Lorentz gauge condition
(\ref{lorentz}).  What is the total ``gravitational force'' on the
particle?

Although the above question is closely analogous to the question posed
at the beginning of Sec~\ref{sec:em}, there still remain a several
notable differences between the electromagnetic and gravitational
cases. First, since we have made a linearized approximation, it is
necessary here that $\gamma^{\rm in}_{ab}$ be ``small'' compared with
the background metric $g^{(0)}_{ab}$. No corresponding restriction on
$F^{\rm in}_{ab}$ was necessary in the electromagnetic case. This
restriction on $\gamma^{\rm in}_{ab}$ will have an important bearing
on the final form of the reduced order equations of motion which we
shall obtain at the end of this section. However, it should be noted
that this restriction on $\gamma^{\rm in}_{ab}$ does not actually
impose any physical restriction on the applicability of our results,
since if we wished to consider a situation where the incoming, free
gravitational radiation is ``large'', we could simply incorporate this
radiation into the background metric $g^{(0)}_{ab}$. Indeed, there
would be no (physical) loss of generality in demanding that
$\gamma^{\rm in}_{ab} = 0$, but we choose not to do so, since there
are a wide variety of circumstances where it is both appropriate and
convenient to treat the incoming radiation as a linearized
perturbation.

The second difference concerns the status of ``external forces''. In
the electromagnetic case, we were free to assume that
$T^{ab}_{\rm{ext}}$ had no coupling to the electromagnetic
field. However, in the gravitational case, it is not consistent to
assume that $T^{ab}_{\rm{ext}}$ has no gravitational coupling; we must
include $T^{ab}_{\rm{ext}}$ on the right side of eq.~(\ref{gravwave}),
and take into account its contributions to $\gamma_{ab}$. Since,
ultimately, we will set $T^{ab}_{\rm{ext}} = 0$ to get the equations
of motion of a freely falling particle, this will not be relevant for
our final formula for the equations of motion. However, in our
expression for $f^a_G$, the presence of $T^{ab}_{\rm{ext}}$ will make
a contribution to $\gamma_{ab}$, which must be included.

A third important difference concerns the gauge invariance of our
results. In the electromagnetic case, both the Maxwell field,
$F_{ab}$, and the world line, $z(\tau)$, of the particle are gauge
invariant. Most importantly, all of the information concerning the
motion of the particle is contained in the specification of
$z(\tau)$. However, in the gravitational case, neither $\gamma_{ab}$
nor $z(\tau)$ are gauge invariant, since both can be changed by
diffeomorphisms. Indeed, $z(\tau)$ can be changed arbitrarily by
diffeomorphisms. Thus, the specification of $z(\tau)$ alone provides
no information about the motion of the particle. Rather, this
information is encoded in the joint specification of both $z(\tau)$
{\em and} $\gamma_{ab}$.

Despite the above differences, our analysis of the gravitational
self-force will now proceed in close parallel with the electromagnetic
case. In order to motivate the axioms which we ultimately will adopt,
we consider a small, nearly spherical extended body with weak
self-gravity, so that the spacetime metric, $g_{ab}$, deviates only
slightly from a vacuum solution, $g^{(0)}_{ab}$. We seek to obtain an
equation of motion for a suitable representative world line in the
body, expressed in terms of the structures associated with the
``background spacetime'' $(M, g^{(0)}_{ab}$). To do so, we view the
exact four-momentum density, $(T_{\rm body}^{ab}) \epsilon_{bcde}$ in
the spacetime $(M, g_{ab})$ from the perspective of the background
spacetime $(M, g^{(0)}_{ab})$. In parallel with eq.~(\ref{defpcs}), at
a point $z(\tau)$ on a representative world line in the body with
tangent $u^a$, we define the four-momentum $p^a$ by
\begin{eqnarray}
k_a p^a &=& \int_{\Sigma^{(0)}} k^{(0)}_a (T_{\rm body}^{ab}) \epsilon_{bcde}.
\end{eqnarray}
Here $\Sigma^{(0)}$ is the hypersurface generated by geodesics of
$g^{(0)}_{ab}$ which are orthogonal (with respect to $g^{(0)}_{ab}$)
to $u^a$, and the vector field~$k^{(0)}_a$ is given the superscript
``0'' in order to emphasize that we are extending $k^a$ off of the
world line by parallel transporting it with respect to the background
metric $g^{(0)}_{ab}$ (as opposed to $g_{ab}$). As in the
electromagnetic case, we assume that a representative world line can
be chosen so that---to an excellent approximation when the body is
sufficiently small and spherical---we have at each point of the world
line
\begin{equation}
p^a = mu^a
\label{pmv2}
\end{equation}
Again, the difficulties in justifying this assumption would provide
one of the more formidable obstacles to converting the motivational
arguments given here into theorems about radiation reaction forces.

From the perspective of the spacetime $(M, g^{(0)}_{ab}$), the force on
the body is given by 
\begin{equation}
f^a \equiv u^b \nabla^{(0)}_b p^a.
\end{equation}
Although we would expect $g_{ab} p^a p^b$ to be constant along the
world line to the order to which we shall work, there is no reason why
$g^{(0)}_{ab} p^a p^b$ need be constant to this order. Equivalently,
if we normalize $u^a$ so that $g^{(0)}_{ab} u^a u^b = -1$, there is no
reason why the parameter $m$ in eq.~(\ref{pmv2}) need be constant
along the curve. If $g^{(0)}_{ab} p^a p^b$ fails to be constant, $f^a$
will fail to be perpendicular to $u^a$ (in the metric $g^{(0)}_{ab}$);
we shall retain the component of $f^a$ parallel to $u^a$ in our
formula for the gravitational force below. However, the deviation from
geodesic motion, $u^b \nabla^{(0)}_b u^a$, is determined entirely by
the projection of $f^a$ perpendicular to $u^a$ in the metric
$g^{(0)}_{ab}$, i.e., we have
\begin{equation}
m u^b \nabla^{(0)}_b u^a = {h^{(0)}}^a{}_b f^b
\label{fproj}
\end{equation}
where $h^{(0)}_{ab} = g^{(0)}_{ab} + u_a u_b$ and all indices here are
raised and lowered using $g^{(0)}_{ab}$. Thus, we ultimately will
project $f^a$ perpendicular to $u^a$ when we wish to obtain the
equation of motion of the particle.

The calculation of $f^a$ proceeds in parallel with the calculation in
the electromagnetic case. Taking $k^a$ to be parallel transported
(with respect to $g^{(0)}_{ab}$) along the world line, we have
\begin{eqnarray}
k_a f^a &=& u^b \nabla^{(0)}_b (k_a p^a) \nonumber \\
        &=& \int_{\Sigma^{(0)}} 
            \pounds_w (k^{(0)}_a T_{\rm body}^{ab} \epsilon_{bcde}).
\end{eqnarray}
Applying the identity (\ref{formsid}) and using Stokes' theorem, we
obtain
\begin{equation}
k_a f^a = \int_{\Sigma^{(0)}} [\nabla_b k^{(0)}_a T_{\rm body}^{ab} 
          + k^{(0)}_a \nabla_b T_{\rm body}^{ab}] w^c \epsilon_{cdef}).
\label{fg}
\end{equation}
In the first term, we rewrite $\nabla_b k^{(0)}_a$ as
\begin{equation}
\nabla_b k^{(0)}_a = \nabla^{(0)}_b k^{(0)}_a - C^c{}_{ba} k_c,
\end{equation}
where
\begin{equation}
C^c{}_{ba} \equiv \frac{1}{2}g^{(0)cd}(\nabla^{(0)}_b \gamma_{ad} +
	\nabla^{(0)}_a \gamma_{bd} - \nabla^{(0)}_d \gamma_{ba})
\label{Cdef}
\end{equation}
Although $\nabla^{(0)}_b k^{(0)}_a$ will make a nonvanishing
contribution to the integrand due to the background curvature, this
contribution is easily seen to vanish in our final expression for
$f^a$ when we take the point particle limit \footnote{Note that
neither $\nabla_b k_a$ nor $\nabla^{(0)}_b k_a$ would be negligible in
the point particle limit if $k^a$ were defined by parallel transport
with respect to $g_{ab}$ rather than $g^{(0)}_{ab}$}, so we will drop
this contribution as well as the other background curvature
corrections mentioned in the electromagnetic derivation as they arise
in the calculations below.

As in the electromagnetic case, for generality, we allow the body to
be coupled to additional classical matter with stress-energy tensor
$T^{ab}_{\rm ext}$.  By conservation of total stress-energy, we have
\begin{equation}
\nabla_b [T_{\rm body}^{ab} +  T^{ab}_{\rm ext}] = 0.
\end{equation}
so that
\begin{equation}
\nabla_b T_{\rm body}^{ab} = - \nabla_b T^{ab}_{\rm ext}.
\end{equation}
Substituting these results in eq.~(\ref{fg}), we have
\begin{equation}
k_a f^a = \int_{\Sigma^{(0)}} 
            - k^{(0)}_a C^a{}_{bc} T_{\rm body}^{bc} w^d d\Sigma_d 
            + k_a f_{\rm ext}^a
\label{fCT}
\end{equation}
where
\begin{equation}
k_a f_{\rm ext}^a \equiv \int_{\Sigma^{(0)}} 
                         - \nabla_b T_{\rm ext}^{ab} w^d d\Sigma_d.
\end{equation}
We now approximate the body to be ``at rest'' at time $\Sigma^{(0)}$,
so that $T_{\rm body}^{bc} = \rho u^b u^c$, where $u^b$ is the unit
normal (in the metric $g^{(0)}_{ab}$) to $\Sigma^{(0)}$. We obtain
\begin{equation}
k_a f_{\rm G}^a \equiv k_a (f^a - f_{\rm ext}^a)
	= \int_{\Sigma^{(0)}} 
            k^{(0)}_a \rho (\frac{1}{2} u^b u^c \nabla^{(0)a} \gamma_{bc}
            - u^b u^c \nabla^{(0)}_b \gamma_c{}^a)
            w^d d\Sigma_d 
\end{equation}
This formula corresponds to eq.~(\ref{fems}) in the electromagnetic
case, with the expression in parentheses playing the role of the
electric field $E^a$ which appeared there. Therefore, we can
immediately write down the gravitational analog of eq.~(\ref{fEM}). We
obtain
\begin{eqnarray}
f_{\rm G}^\alpha &=& 
	m \langle E_{\rm G}^\alpha N\rangle_R
	- \int_{\Sigma(\tau)} \frac{m(r)}{4\pi r^2} 
	\left[\frac{\partial E_{\rm G}^\alpha}{\partial r} N + 
	\hat{r}^\beta a_\beta E_{\rm G}^\alpha\right] \,dV
\nonumber \\
	&& + [\mbox{terms which vanish as $R \rightarrow 0$}]
\label{fgrav}
\end{eqnarray}
where
\begin{equation}
E_{\rm G}^\alpha \equiv \frac{1}{2} u^\beta u^\gamma 
	\nabla^\alpha \gamma_{\beta\gamma} - u^\beta u^\gamma 
	\nabla_\beta \gamma_\gamma{}^\alpha.
\end{equation}
In this equation and in all equations henceforth, it is to be
understood that all quantities except $\gamma_{ab}$ refer to the
background structure, and the superscript ``0'' will be omitted on the
background metric and its derivative operator.  As in the
electromagnetic case, a ``point particle'' limit of the right side of
eq.~(\ref{fgrav}) cannot be taken in a straightforward
manner. However, we can again consider the {\em difference} in $f_{\rm
G}^\alpha$ on two bodies of similar composition that move on different
world lines in (possibly different) background spacetimes. In order to
find the conditions under which such a difference will remain bounded
as $R \rightarrow 0$, we once again study the singular behavior of the
exterior field of a point source in curved spacetime using the
Hadamard expansion techniques of Dewitt and Brehme \cite{db}. Since
the trace-reversed metric perturbation $\bar{\gamma}_{ab}$ satisfies a
wave equation (\ref{gravwave}) very similar to the equation for the
electromagnetic vector potential, the Hadamard expansion goes through
in close parallel with the electromagnetic case. The covariant
expansions for $\bar{\gamma}^\pm_{a'b'}$ and
$\nabla_{c'} \bar{\gamma}^\pm_{a'b'}$ (the gravitational
analogs of eq.~\ref{dbeqn}) are given by
\begin{eqnarray}
\bar{\gamma}^\pm_{a'b'} (x)
&=& 2m \bar{g}_{a'(a} \bar{g}_{|b'|b)} [
2 r^{-1} u^a u^b
\pm 4 \kappa^{-2} a^a u^b]
\nonumber \\ &&
\pm m \int^{\pm\infty}_{\tau^\pm}
	G^\pm_{a'b'a''b''}
        u^{a''}(\tau'') u^{b''}(\tau'') \, d\tau''
+ O(r)
\label{gamma}
\end{eqnarray}
and
\begin{eqnarray}
\nabla_{c'} \bar{\gamma}^\pm_{a'b'} (x)
&=& 2m \bar{g}_{c'c} \bar{g}_{a'(a} \bar{g}_{|b'|b)} [
- 2 r^{-2} \kappa^{-1} u^{a} u^{b} \Omega^{c}
- 4 r^{-1} \kappa^{-3} a^{a}u^{b} u^{c}
- r^{-1} \kappa^{-3} u^{a} u^{b} a^{c}
\nonumber \\ &&
- \frac{1}{4} \kappa^{-5} a^2 u^{a} u^{b} \Omega^{c}
+ \kappa^{-5} u^{a} u^{b} u^{c} \dot{a}^d \Omega_d
+ 2 \kappa^{-3} \dot{a}^{a} u^{b} \Omega^{c}
+ 2 \kappa^{-3} a^{a} a^{b} \Omega^{c}
\nonumber \\ &&
\pm \frac{2}{3} \kappa^{-6} a^2 u^{a} u^{b} u^{c}
\mp 4 \kappa^{-4} \dot{a}^{a} u^{b} u^{c}
\mp 4 \kappa^{-4} a^{a} a^{b} u^{c}
\mp 4 \kappa^{-4} a^{a} u^{b} a^{c}
\mp \frac{2}{3} \kappa^{-4} u^{a} u^{b} \dot{a}^{c}
\nonumber \\ &&
- \frac{2}{3} \kappa^{-3} R_{def}{}^c u^{a} u^{b} u^d \Omega^e u^f
- 2 \kappa ^{-3} R_{def}{}^a u^{b} u^{c} u^d \Omega^e u^f
+ 2\kappa^{-1} R^c{}_{ed}{}^a u^{b} u^d \Omega^e
\nonumber \\ &&
- 2 \kappa^{-1} R_d{}^a{}_e{}^b \Omega^{c} u^d u^e
\mp 2 \kappa^{-2} R^c{}_{de}{}^a u^{b} u^d u^e
\pm 2 \kappa^{-2} R_d{}^a{}_e{}^b u^{c} u^d u^e]
\nonumber \\ &&
\pm m \int^{\pm\infty}_{\tau^\pm} \nabla_{c'}
	G^\pm_{a'b'a''b''}
        u^{a''}(\tau'') u^{b''}(\tau'') \, d\tau''
+ O(r)
\label{gradgamma}
\end{eqnarray}
(In these formulas, we have normalized the advanced and retarded
Green's functions $G^\pm_{aba'b'}$ so that they satisfy
eq.~(\ref{gravwave}) with source $-16 \pi \bar{g}_{aa'} \bar{g}_{bb'}
\delta(x,z)$.)

Expanding $\bar{g}_{\alpha'\alpha}$ and $\kappa$, we find that the
Riemann normal coordinate components of $\nabla_c
\bar{\gamma}^\pm_{a'b'}$, in the same notation as
eq.~(\ref{Fexpand}), are
\begin{eqnarray}
\nabla_{\gamma'} \bar{\gamma}^\pm_{\alpha'\beta'} (x) &=& 2m [
-2 r^{-2} u_{\alpha'} u_{\beta'} \Omega_{\gamma'}
-4 r^{-1} a_{(\alpha'}u_{\beta')} u_{\gamma'}
- r^{-1} u_{\alpha'} u_{\beta'} a_{\gamma'}
+ r^{-1} u_{\alpha'} u_{\beta'} \Omega_{\gamma'}
	a^\delta \Omega_\delta
\nonumber \\ &&
+ 6 a_{(\alpha'} u_{\beta')} u_{\gamma'} a^\delta \Omega_\delta
+ \frac{3}{2} u_{\alpha'} u_{\beta'} a_{\gamma'} (a^\delta \Omega_\delta)
- \frac{3}{4} u_{\alpha'} u_{\beta'} \Omega_{\gamma'}
	(a^\delta \Omega_\delta)^2
+ u_{\alpha'} u_{\beta'} u_{\gamma'} \dot{a}^\delta \Omega_\delta
\nonumber \\ &&
+ \frac{1}{3} u_{\alpha'} u_{\beta'} \Omega_{\gamma'}
	R_{\delta\lambda\epsilon\kappa}
	u_{\delta} u^\epsilon \Omega^\lambda \Omega^\kappa
- \frac{2}{3} R_{\alpha'\delta\lambda\epsilon} u^\lambda \Omega^\delta
  	\Omega^\epsilon u_{\beta'} \Omega_{\gamma'}
- \frac{1}{4} a^2 u_{\alpha'} u_{\beta'} \Omega_{\gamma'}
\nonumber \\ &&
+ 2 \dot{a}_{(\alpha'} u_{\beta')} \Omega_{\gamma'}
+ 2 a_{\alpha'} a_{\beta'} \Omega_{\gamma'}
\pm \frac{2}{3} a^2 u_{\alpha'} u_{\beta'} u_{\gamma'}
\mp 4 \dot{a}_{(\alpha'} u_{\beta')} u_{\gamma'}
\mp 4 a_{\alpha'} a_{\beta'} u_{\gamma'}
\nonumber \\ &&
\mp 4 a_{(\alpha'} u_{\beta')} a_{\gamma'}
\mp \frac{2}{3} u_{\alpha'} u_{\beta'} \dot{a}_{\gamma'}
- \frac{2}{3} R_{\delta\epsilon\lambda\gamma'} u_{\alpha'} u_{\beta'} u^\delta
	\Omega^\epsilon u^\lambda
- 2 R_{\delta\epsilon\lambda(\alpha'} u_{\beta')} u_{\gamma'} u^\delta
	\Omega^\epsilon u^\lambda
\nonumber \\ &&
+ 2 R_{\gamma'\epsilon\delta(\alpha'}u_{\beta')} u^\delta \Omega^\epsilon
- 2 R_{\delta\alpha'\epsilon\beta'} \Omega_{\gamma'} u^\delta u^\epsilon
\mp 2 R_{\gamma'\delta\epsilon(\alpha'} u_{\beta')} u^\delta u^\epsilon
\pm 2 R_{\delta\alpha'\epsilon\beta'}
	u_{\gamma'} u^\delta u^\epsilon]
\nonumber \\ &&
\pm m \int^{\pm\infty}_{\tau^\pm} \nabla_{\gamma'}
	G^\pm_{\alpha'\beta'\alpha''\beta''}
        u^{\alpha''}(\tau'') u^{\beta''}(\tau'') \, d\tau''
+ O(r)
\label{gradgammaexp}
\end{eqnarray}

We have verified (\ref{gamma}), (\ref{gradgamma}), and
(\ref{gradgammaexp}) using the software package MathTensor. There are
significant discrepancies between eqs.~(\ref{gamma}) and
(\ref{gradgamma}) and the corresponding equations in
\cite{mino}. Specifically, (i) there is a sign discrepancy between the
term $\pm 4 \kappa^{-2} a^a u^b$ in eq.~(\ref{gamma}) and the
corresponding term in eq.~(2.27) of \cite{mino}, (ii) many of the
terms in eq.~(\ref{gradgamma}) which do not involve the Riemann
curvature fail to appear in eq.~(2.33) of \cite{mino}, and (iii) there
are several sign discrepancies between the Riemann curvature terms of
eq.~(\ref{gradgamma}) and eq.~(2.33) of \cite{mino}. The sign
discrepancies in the terms involving the Riemann curvature appear to
arise from the use by \cite{mino} of inconsistent sign conventions in
various formulas (the conventions of \cite{wald} and \cite{db} both
seem to appear), but we have not attempted to track down the origin of
the other discrepancies.

Aside from the obvious complexity introduced by the additional index
structure, there are two important differences between the above
formulas and the corresponding formulas in the electromagnetic
case. First, as stated above, we have assumed here that the background
spacetime is a solution of the vacuum Einstein equation, so in the
gravitational case, no terms are present which involve the Ricci
curvature. Although it would be possible to repeat the above analysis
by perturbing off of a nonvacuum solution, the perturbations of the
metric and background matter would become coupled at linear order, so
eq.~(\ref{gravwave}) no longer would hold, and the entire analysis
would have to be redone.  Second, due to the presence of several terms
of alternating sign in the above expressions which do not depend upon
curvature, we see that the advanced and retarded expressions for
$\gamma_{ab}$ and its first spatial derivative for a uniformly
accelerating trajectory in flat spacetime do not agree in a
neighborhood of the worldline of the particle. In parallel with the
electromagnetic case, the advanced and retarded solutions can be shown
to be gauge equivalent within the entire ``Rindler wedge'' containing
the worldline.  However, unlike the electromagnetic case, the analog
of the Lorentz force, $-C^a_{bc} u^b u^c$, is not gauge invariant in
this case, and it differs for the advanced and retarded solutions even
in the limit as $r \rightarrow 0$. Nevertheless, it can be verified
that this difference between the forces for the advanced and retarded
solutions is parallel to the four-velocity of the particle. Therefore,
when we project the force perpendicular to the particle's
four-velocity to produce an equation of motion, the difference will
vanish and the situation is effectively the same as in the
electromagnetic case.

Despite the above differences, eq.~(\ref{gradgammaexp}) shares the
most important property of the analogous electromagnetic expression
(\ref{Fexpand}), namely the divergent terms as $r \rightarrow 0$
depend only upon the four-velocity and four-acceleration of the
particle at $z(\tau)$. Therefore, in direct analogy with Axiom
\ref{ax:comp} in the electromagnetic case. we postulate the following

\begin{gaxiom}[Comparison axiom] \label{ax:compgrav}
Consider two points,~$P$ and~$\widetilde{P}$, each lying on timelike
world lines in (possibly different) spacetimes which contain
linearized metric perturbations sourced by particles of mass $m$ on
the world lines (see eq.~(\ref{gravwave})). If the four-accelerations
of the world lines at~$P$ and~$\widetilde{P}$ have the same magnitude,
and if we identify the neighborhoods of~$P$ and~$\widetilde{P}$ via
Riemann normal coordinates such that the four-velocities and
four-accelerations are identifed, then the difference in the
gravitational forces~$f_{\rm G}^a$ and~$\tilde{f}_{\rm G}^a$, is given
by the limit as~$r \rightarrow 0$ of the difference of the effective
gravitational forces averaged over a sphere at geodesic distance~$r$
from the world line at $P$.
\begin{equation}
f_{\rm G}^a - \tilde{f}_{\rm G}^a = \lim_{r \rightarrow 0} \left(
	\left\langle (\frac{1}{2} \nabla^a \gamma_{bc} 
	- \nabla_b \gamma_c{}^a)
	- (\frac{1}{2} \nabla^a \widetilde{\gamma}_{bc} 
	- \nabla_b \widetilde{\gamma}_c{}^a) \right\rangle
	\right) u^b u^c
\end{equation}
\end{gaxiom}

In analogy with Axiom \ref{ax:flat}, we also postulate

\begin{gaxiom}[Flat spacetime axiom] \label{ax:flatgrav}
If~$(M,g_{ab})$ is Minkowski spacetime, the world line is uniformly
accelerating, and~$\gamma_{ab}$ is the half-advanced, half-retarded
solution, $\gamma_{ab} =\frac{1}{2}[\gamma^-_{ab} + \gamma^+_{ab}]$,
then~$f_{\rm G}^a=0$ at every point on the world line.
\end{gaxiom}
As noted above, since $-C^a_{bc} u^b u^c$ differs for the advanced and
retarded solutions, it does matter in this case that we use the
half-advanced, half-retarded solution in this axiom, rather than, say,
the advanced or retarded solution, although this difference does not
affect the projection of the force perpendicular to the worldline of
the particle.

In parallel with the electromagnetic case, the above axioms yield the
following prescription for the gravitational force. Let $(M,g_{ab})$
be a solution of the vacuum Einstein equation and let $\gamma_{ab}$ be
a solution of eq.~(\ref{gravwave}). At a point $P$ on the particle's
world line, we compare $\gamma_{ab}$ with the half-advanced,
half-retarded solution for a uniformly accelerating trajectory in the
tangent space (using the exponential map to make the comparison).  The
gravitational force is then given by calculating the difference in
$-C^a_{bc} u^b u^c= (\frac{1}{2} \nabla^a \gamma_{bc} - \nabla_b
\gamma_c{}^a)u^b u^c$ for these two fields, averaging over a sphere of
radius $r$, and letting $r \rightarrow 0$. If we write $\gamma_{ab}$
as $\gamma_{ab} = \gamma^{\rm in}_{ab} + \gamma^-_{ab}$, the resulting
expression is
\begin{eqnarray}
f_{\rm G}^a &=& m(\frac{1}{2} \nabla^a \gamma^{\rm in}_{bc}
	- \nabla_b (\gamma^{\rm in})_c{}^a) u^b u^c
	- m^2 (\frac{11}{3} \dot{a}^a + \frac{1}{3} a^2 u^a)
\nonumber \\
	&& + m^2 u^b u^c \int_{-\infty}^{\tau^-} 
	(\frac{1}{2} \nabla^a G^-_{bca'b'} - \nabla_b (G^-)_c{}^a{}_{a'b'})
	u^{a'} u^{b'}
	\,d\tau'
\label{fgravfinal}
\end{eqnarray}

As anticipated, this expression contains contributions to $f_{\rm
G}^a$ parallel to the four-velocity $u^a$, which merely describe the
effect of the metric perturbation on the normalization of $u^a$. To
obtain the equations of motion, we project $f_{\rm G}^a$ perpendicular
to $u^a$ as in Eq.~(\ref{fproj}) above. This yields
\begin{eqnarray}
a^a &=& (\frac{1}{2} \nabla^a \gamma^{\rm in}_{bc}
	- \nabla_b (\gamma^{\rm in})_c{}^a 
	- \frac{1}{2} u^a u^d \nabla_d \gamma^{\rm in}_{bc}) u^b u^c
	- \frac{11}{3} m (\dot{a}^a - a^2 u^a)
\nonumber \\
&& + m u^b u^c \int_{-\infty}^{\tau^-} 
	(\frac{1}{2} \nabla^a G^-_{bca'b'} - \nabla_b (G^-)_c{}^a{}_{a'b'}
 	- \frac{1}{2} u^a u^d \nabla_d G^-_{bca'b'})
	u^{a'} u^{b'}
\,d\tau'
\label{eqmotiongrav}
\end{eqnarray}

It should be noted that an Abraham-Lorentz term of the form
$(\dot{a}^a - a^2 u^a)$ appears in eq.~(\ref{eqmotiongrav}), but with
a sign opposite to that of the electromagnetic case, corresponding to
the ``antidamping'' phenomenon found by Havas \cite{havas}.

Finally, we apply to this equation of motion the same reduction of
order techniques that we applied to the electromagnetic equation of
motion. When we do so, the terms on the right side involving
$m\dot{a}^a$ and $ma^a$ get eliminated in favor of terms involving $m
\gamma^{\rm in}_{ab}$. However, our equation (\ref{eqmotiongrav}) is
valid only to linear order in both $m$ and $\gamma^{\rm in}_{ab}$, so
it is not consistent to keep terms involving products of these two
quantities. (This contrasts strongly with the electromagnetic case, we
we worked only to linear order in $e^2/m$, but $F^{\rm in}_{ab}$ was
allowed to be as large as we liked, so that $ (e/m) F^{\rm in}_{ab}$
could be treated as being of order unity.)  Consequently, the
reduction of order procedure in this case effectively drops the
Abraham-Lorentz terms, leaving the ``tail term'' as the only
contribution to the ``self-force''\footnote{This contrasts sharply
with the analysis of \cite{havas} which effectively neglected the
dominant ``tail term''.}:
\begin{eqnarray}
a^a &=& (\frac{1}{2} \nabla^a \gamma^{\rm in}_{bc}
	- \nabla_b (\gamma^{\rm in})_c{}^a 
	- \frac{1}{2} u^a u^d \nabla_d \gamma^{\rm in}_{bc}) u^b u^c
\nonumber \\
	&& + m u^b u^c \int_{-\infty}^{\tau^-} 
	(\frac{1}{2} \nabla^a G^-_{bca'b'} - \nabla_b (G^-)_c{}^a{}_{a'b'}
 	- \frac{1}{2} u^a u^d \nabla_d G^-_{bca'b'})	
	u^{a'} u^{b'}
	\,d\tau'
\label{gravreduct}
\end{eqnarray}
This formula agrees with the results of Mino et al. \cite{mino},
although they did not include $\gamma^{\rm in}_{ab}$ in their
expression. Not that eq.~(\ref{gravreduct}) has a very simple
interpretation: To lowest nontrivial order, the particle moves on a
geodesic of $g^{(0)}_{ab} + \gamma_{ab}$, where $\gamma_{ab} =
\gamma^{\rm in}_{ab} + \gamma^{\rm tail}_{ab}$ and $\bar{\gamma}^{\rm
tail}_{ab} = \bar{\gamma}^{\rm tail}_{ab} - \frac{1}{2} g^{(0)}_{ab}
\gamma^{\rm tail}$ is the last term in equation (\ref{gamma}).

\section{Conclusions}
\label{conclusions}

In this paper, we have taken an axiomatic approach to obtain the
lowest order electromagnetic and gravitational ``self-forces'' on a
small, nearly spherical body of sufficiently small charge and/or
mass. Our final result for the total electromagnetic force on a body
(possibly acted upon by an external electromagnetic field as well as
additional, non-electromagnetic ``external forces'') is given by
eq.~(\ref{eqmotion}). If such a body is subject only to
electromagnetic forces, our final result for the (reduced order)
equation of motion of the body is given by eq.~(\ref{reduct}). The
corresponding results for the gravitational case are given by
eqs.~(\ref{fgravfinal}) and (\ref{gravreduct}).

The above electromagnetic results were derived from the two axioms
given in Sec.~\ref{sec:emax}, and the corresponding gravitational
results were obtained from the axioms of Sec.~\ref{sec:gr}. Although
plausibility arguments in support of these axioms were given, we did
not attempt to prove that the electromagnetic axioms follow as a
consequence of Maxwell's equations in curved spacetime together with
conservation of total stress-energy, nor did we attempt to prove that
the gravitational axioms follow from Einstein's
equation. Nevertheless, we believe that our plausibility arguments
have provided the first steps in that direction. In any case, the
problem of providing a rigorous justification for the electromagnetic
and gravitational self-forces and the corresponding equations of
motion has been reduced to the problem of providing a rigorous
justification of our axioms.

Finally, we note that on account of the ``tail term'', our equations
of motion in both the electromagnetic and gravitational cases are
integro-differential equations which, in principle, require us to know
the entire past history of the particle. However, if the curvature of
spacetime is sufficiently small and the motion of the body is
sufficiently ``slow'', one would expect the ``tail term'' to become
effectively local, since the contributions to the ``tail term''
arising from portions of the orbit distant from the present position
of the particle should become negligible. Indeed, if the tail term
becomes effectively local, arguments using the standard dipole formula
for radiated energy together with conservation of energy (see
\cite{jackson} and the analogous gravitational calculation given
below) suggest that if $F^{\rm in}_{ab} = 0$ and $R_{ab} = 0$, the
``tail term'' of eq.~(\ref{reduct}) should reduce to the familiar
Abraham-Lorentz damping force
\begin{equation}
\label{al}
f_{\rm EM}^i = \frac{2}{3} e^2 \frac{d^3 x^i}{dt^3}.
\end{equation}
That this is indeed the case was established by DeWitt and DeWitt
\cite{dd} for a charged particle in a slow, circular orbit in
linearized Schwarzschild spacetime. It should bve emphasized that the
``true'' Abraham-Lorentz force actually vanishes here---since
$a^a=\dot{a}^a=0$ for geodesic motion--- but, remarkably, the tail
term mocks up an effective Abraham-Lorentz term.

In the gravitational case, the standard ``quadrupole formula'' for
radiated power in the slow motion, weak field limit is
\begin{equation}
\label{quad}
P = \frac{1}{45} \frac{d^3 Q^{ij}}{dt^3} \frac{d^3 Q_{ij}}{dt^3}
\end{equation}
where $i$ and $j$ are spatial indices ($i,j=1,2,3$) for a set of
global inertial coordinates and $Q^{ij}$ is the traceless quadrupole
moment
\begin{eqnarray}
Q^{ij} &\equiv& q^{ij} - \frac{1}{3} q \delta^{ij} \\
q^{ij} &\equiv& 3 \int T^{00} x^i x^j \, d^3 x.
\end{eqnarray}
so, for a point particle,
\begin{equation}
\label{pointQ}
Q^{ij} = 3 m \left(x^i x^j - \frac{1}{3}x^k x_k \delta^{ij}\right).
\end{equation}
By conservation of energy, we should have
\begin{eqnarray}
\int f_{\rm G}^i v_i \, dt 
	&=& -\frac{1}{45} \int \frac{d^3 Q^{ij}}{dt^3} 
            \frac{d^3 Q_{ij}}{dt^3} \,dt
\nonumber \\
	&=& -\frac{1}{45} \int \frac{d^5 Q^{ij}}{dt^5} 
            \frac{dQ_{ij}}{dt} \,dt
\nonumber \\
        &=& -\frac{6}{45} m\int \frac{d^5 Q^{ij}}{dt^5} 
            \left(v_i x_j - \frac{1}{3} v^k v_k \delta_{ij}\right) \,dt
\nonumber \\
        &=& -\frac{2}{15} m \int \left(\frac{d^5 Q^{ij}}{dt^5} x_j\right)
                           v_i \, dt.
\label{intparts}
\end{eqnarray}
which suggests that the radiation reaction force should be given by
(see section 36.8 of Misner, Thorne, and Wheeler \cite{wald})
\begin{equation}
\label{fquad}
f_{\rm G}^i = -\frac{2}{15} m \frac{d^5 Q^{ij}}{dt^5} x_j.
\end{equation}
It would be interesting to perform the gravitational analog of the
analysis of DeWitt and DeWitt \cite{dd} to see if this formula does,
indeed, arise from the tail term of eq.~(\ref{gravreduct}) in the slow
motion, weak field limit.

\section*{Acknowledgements}

We wish to thank Eanna Flanagan for reading the manuscript. This
research was supported in part by NSF grant PHY 95-14726.


\begin{thebibliography}{99}

\bibitem{gt} R. Geroch and J. Traschen, Phys. Rev. D {\bf 36}, 1017 (1987).

\bibitem{gj} R. Geroch and P. S. Jang, J. Math. Phys {\bf 16}, 65
(1975).

\bibitem{pen} R. Penrose and W. Rindler, {\em Spinors and Space-time,
vol. I} (Cambridge University Press, Cambridge, 1984). W.G. Unruh,
Proc. Roy. Soc. Lond. A{\bf 348}, 447 (1976).

\bibitem{dixon74} W. G. Dixon, Trans. Roy. Soc. Lond. A. {\bf 277}, 59
(1974).

\bibitem{db} B. S. DeWitt and R. W. Brehme, Annals of Physics {\bf 9},
220 (1960).

\bibitem{hobbs} J. M. Hobbs, Ann. Phys. {\bf 47}, 141 (1968).

\bibitem{mino} Y. Mino, M. Sasaki, and T. Tanaka, preprint
gr-qc/9606018.

\bibitem{beig} W. Beiglb\"{o}ck, Commun. Math. Phys. {\bf 5}, 106 (1967).

\bibitem{dixon64} W. G. Dixon, Nuovo Cim. {\bf 34}, 319 (1964).

\bibitem{hormander} L. H\"{o}rmander, {\em The Analysis of Linear
Partial Differential Operators IV} (Springer, Berlin, 1985)

\bibitem{wald} C.W. Misner, K.S. Thorne, and J. A. Wheeler, {\em
Gravitation} (Freeman, San Francisco, 1970). R. M. Wald, {\em General
Relativity} (University of Chicago Press, Chicago, 1984).

\bibitem{jackson} J. D. Jackson, {\em Classical Electrodynamics}
(Wiley, New York, 1975).

\bibitem{boulware} D. Boulware, Ann. Phys. {\bf 124}, 169 (1980).

\bibitem{hartle} J. B. Hartle and K. S. Thorne, Phys. Rev. D {\bf 31},
1815 (1985).

\bibitem{eih} A. Einstein, L. Infeld, and B. Hoffmann, Ann. Math. {\bf
39}, 65 (1938).

\bibitem{flanwald} E. E. Flanagan and R. M. Wald, Phys Rev. D (in
press), preprint gr-qc/9602052.

\bibitem{havas} P. Havas, Phys. Rev. {\bf 108}, 1351 (1957). (A
crucial sign error in eq.~(10) of the reference was corrected in
P. Havas and J. N. Goldberg, Phys. Rev. {\bf 128}, 398 (1962).)

\bibitem{dd} B. S. DeWitt and C. M. DeWitt, Physics {\bf 1}, 3 (1964).

\end{thebibliography}
\end{document}